\RequirePackage{fix-cm}
\documentclass[smallextended]{svjour3}       
\smartqed  
\usepackage{graphicx}
\usepackage{url}
\usepackage{hyperref}
\usepackage{natbib}
\usepackage{rotating}
\usepackage{booktabs}
\usepackage{graphicx}
\usepackage[linesnumbered, ruled, vlined]{algorithm2e}
\usepackage{mathtools}
\DeclarePairedDelimiter\ceil{\lceil}{\rceil}
\begin{document}

\title{From Protocol to Screening: A Hybrid Learning Approach for Technology-Assisted Systematic Literature Reviews\thanks{This research  is co-financed by Greece and the European Union (European Social Fund- ESF)  through the Operational Programme Human Resources Development, Education and  Lifelong Learning in the context of the project Strengthening Human Resources Research Potential via Doctorate Research (MIS-5000432), implemented by the State Scholarships Foundation (IKY).}
}

\titlerunning{A Hybrid Learning Approach for Technology-Assisted SLRs}        






\author{Athanasios Lagopoulos \and Grigorios Tsoumakas 
}


\institute{A. Lagopoulos \and G.Tsoumakas \at
     Aristotle University of Thessaloniki, Thessaloniki, Greece \\ 
     \email{lathanag@csd.auth.gr, greg@csd.auth.gr}     
}

\date{Received: date / Accepted: date}

\maketitle

\begin{abstract}
In the medical domain, a Systematic Literature Review (SLR) attempts to collect all empirical evidence, that fit pre-specified eligibility criteria, in order to answer a specific research question. The process of preparing an SLR consists of multiple tasks that are labor-intensive and time-consuming, involving large monetary costs. Technology-assisted review (TAR) methods  automate the different processes of creating an SLR and they are particularly focused on reducing the burden of screening for reviewers. We present a novel method for TAR that implements a full pipeline from the research protocol to the screening of the relevant papers. Our pipeline overcomes the need of a Boolean query constructed by specialists and consists of three different components: the primary retrieval engine, the inter-review ranker and the intra-review ranker, combining learning-to-rank techniques with a relevance feedback method. In addition, we contribute an updated version of the Task 2 of the CLEF 2019 eHealth Lab dataset, which we make publicly available. Empirical results on this dataset show that our approach can achieve state-of-the-art results.   

\end{abstract}

\section{Introduction}

Systematic literature reviews (SLRs) in the medical domain seek to collect evidence from research publications that fit strict pre-specified eligibility criteria to answer a specific research question. They aim to minimize bias by using explicit, systematic methods documented in advance with a protocol \citep{higgins2019cochrane}. Clinicians practice evidence-based medicine (EBM) by integrating their expertise with the best available external clinical evidence from SLRs \citep{sackett1997evidence}. 

The process of preparing an SLR consists of multiple tasks that can be organized into four general stages: {\em preparation}, {\em retrieval}, {\em appraisal}, and {\em synthesis} \citep{tsafnat2014systematic}. The preparation stage includes the tasks of developing the research question, searching for relevant SLRs, writing the needed protocol, and defining a search strategy. The search strategy includes Boolean queries adapted for each medical database. These Boolean queries typically have very complicated syntax and are usually built by highly trained information specialists. The queries are submitted to medical databases during the retrieval stage, resulting in a vast set of possibly relevant studies \citep{Kanoulas2018}. Subsequently, every study is screened in the appraisal stage, using the title and abstract (abstract-level assessment), and irrelevant studies are removed. Additional assessment is conducted based on the full-text (content-level assessment) of the remaining studies. In the last stage, data are extracted, converted, and synthesized from the relevant studies. The final scientific paper incorporates all this data with the addition of a meta-analysis of the included studies.

Current methods of creating an SLR are labor-intensive and time-consuming, involving large monetary costs. On average, an SLR costs approximately more than \$140K and the time a scientist spends to complete it is 1.72 years \citep{Michelson2019}. The most cost-intensive and time-consuming part when creating a systematic review is the screening process, i.e. the appraisal stage. On average, more than 5,500 documents are returned from the databases and less than 4.7\% (3\%) of them are relevant at the abstract (document) level \citep{borah2017analysis,Kanoulas2018}. Technology-assisted review (TAR) is a relatively new computer science field that employs information retrieval, machine learning and natural language processing (NLP) techniques, which are usually combined with domain specific knowledge, to reduce the workload on screening for SLRs \citep{miwa2014reducing}.

The primary contribution of this work is a novel method for TAR with state-of-the-art results. Our method implements a full pipeline from protocol to screening papers with state-of-the-art results, assisting the researcher in three parts of the process of preparing an SLR: the preparation, the retrieval, and the appraisal. The key novel characteristics of our method are: 
\begin{itemize}
    \item it relies solely on the protocol of a systematic review, overcoming the need for constructing a specialized Boolean query (preparation).
    \item it incorporates domain specialized features deriving from the latest advances of the NLP field (retrieval and appraisal).
\end{itemize}

As a secondary contribution, we publish an updated version of the Task 2 of the CLEF 2019 eHealth lab dataset \citep{kanoulas2019clef}. The updated  dataset fixes previous format issues and provides an up-to-date version of the dataset that includes the latest revisions of the included SLRs. 

This work is an extension of a previously published conference paper \citep{lagopoulos2018learning}. Specifically, we extended our previous pipeline with a primary retrieval engine, fine-tuned the inter-review ranker features and adopted sentence embeddings for both inter-review and intra-review rankers. Our approach was one of the top approaches in Task 2 of the eHealth Lab of CLEF 2017 \citep{anagnostou2017combining} and CLEF 2018 \citep{minas2018aristotle}.

The rest of this paper is organized as follows:  After  a  discussion  of  the related work in Section \ref{sec:related}, we introduce our approach in Section \ref{sec:methodology}.  In Section \ref{sec:empirical}, we describe our case study and the  corresponding  dataset, and then discuss the results of our study, including a comparison of our method with the state of the art. Finally, in Section \ref{sec:conclusion}, we conclude this work and draw future directions.

\section{Related Work}
\label{sec:related}

Several research papers have been published in the past on applications of text mining to assist in identifying relevant studies for a systematic review. Most of them are focused on reducing the number of studies needed to screen, increasing the speed of screening, and improving the workflow through screening prioritization. Early studies were focused on classifying documents as relevant or not to a review topic \citep{cohen2006reducing} and dealing with imbalanced datasets \citep{cohen2006effective}. Later studies evaluated active learning techniques to deal with class imbalance \citep{wallace2010active,wallace2010semi,miwa2014reducing} and several studies exploited the advantages of using the Naive Bayes algorithm \citep{bekhuis2012screening, matwin2010new, frunza2010building,frunza2011exploiting}. Furthermore, researchers experimented with algorithms such as EvoSVM \citep{bekhuis2010towards} and k-nearest neighbors \citep{miwa2014reducing} and different representations such as visual data mining \citep{felizardo2013use} and Latent Dirichlet Allocation (LDA) \citep{miwa2014reducing}. Finally, \cite{cohen2009cross} introduced an approach, combining topic-specific training data with data from other SLR topics and \cite{karimi2010boolean} were the first to compare ranked retrieval with Boolean querying.

The wide variety of datasets, algorithms, and evaluation methods explored in the above studies makes it difficult to draw any conclusions about the most effective approach. To address this issue, the CLEF eHealth Lab organized a task on Technology-Assisted Reviews in Empirical Medicine \citep{kanoulas2017clef,Kanoulas2018, kanoulas2019clef} from 2017 to 2019. The task was aiming to bring together academic, commercial, and government researchers conducting experiments and sharing results on automatic methods to retrieve relevant studies. Lab participants were provided with a set of systematic review topics that were constructed by Cochrane\footnote{\url{https://www.cochrane.org/}} experts. Each topic contained the title and protocol of a systematic review and the corresponding Boolean query. The task was divided into two sub-tasks. In the ``No Boolean Query" sub-task, participants had to complete the search effectively and efficiently bypassing the construction of the Boolean query. Therefore, participants had to first retrieve the documents from PubMed. In the second sub-task, called ``Abstract and Title Screening", participants had to produce an efficient ordering of the documents, such that all of the relevant abstracts are retrieved as early as possible. The set of documents returned from the submitted Boolean query were also provided in this case.

There was a great interest in this task, with many teams participating and presenting different and specialized approaches. Most of the participants proposed different learning-to-rank approaches \citep{Hollmann2017,Scells2017,Chen2017}, while others also adopted active learning \citep{Cormack2017,Cormack2018,Yu2017} and sampling techniques \citep{DiNunzio2017,DiNunzio2018,Nunzio2019,Li2019}. Two teams worked with neural networks and deep learning \citep{Singh2017,Lee2017}. Furthermore, participants represented the textual data in a variety of ways, including topic models \citep{VanAltena2017,Kalphov2017}, TF-IDF \citep{Alharbi2017,Alharbi2018,Alharbi2019}, n-grams \citep{Norman2017,Norman2018,Cohen2018} and text embeddings \citep{Hollmann2017,Chen2017,Wu2018}.

Recently, \cite{Zou2018, Zou2020} proposed an approach that looks for entities in the documents and asks questions to the users to retrieve the last relevant documents. Also, \cite{Scells2020} proposed a computational approach to objectively derive search strategies for systematic reviews and also presented a novel approach that ranks documents for systematic review literature using rank fusion applied to coordination level matching by taking advantage of the boolean query \citep{Scells2020OldDog}.

\section{Our Hybrid Learning Approach}
\label{sec:methodology}
\begin{figure}[ht]
    \centering
    \includegraphics[width=\textwidth]{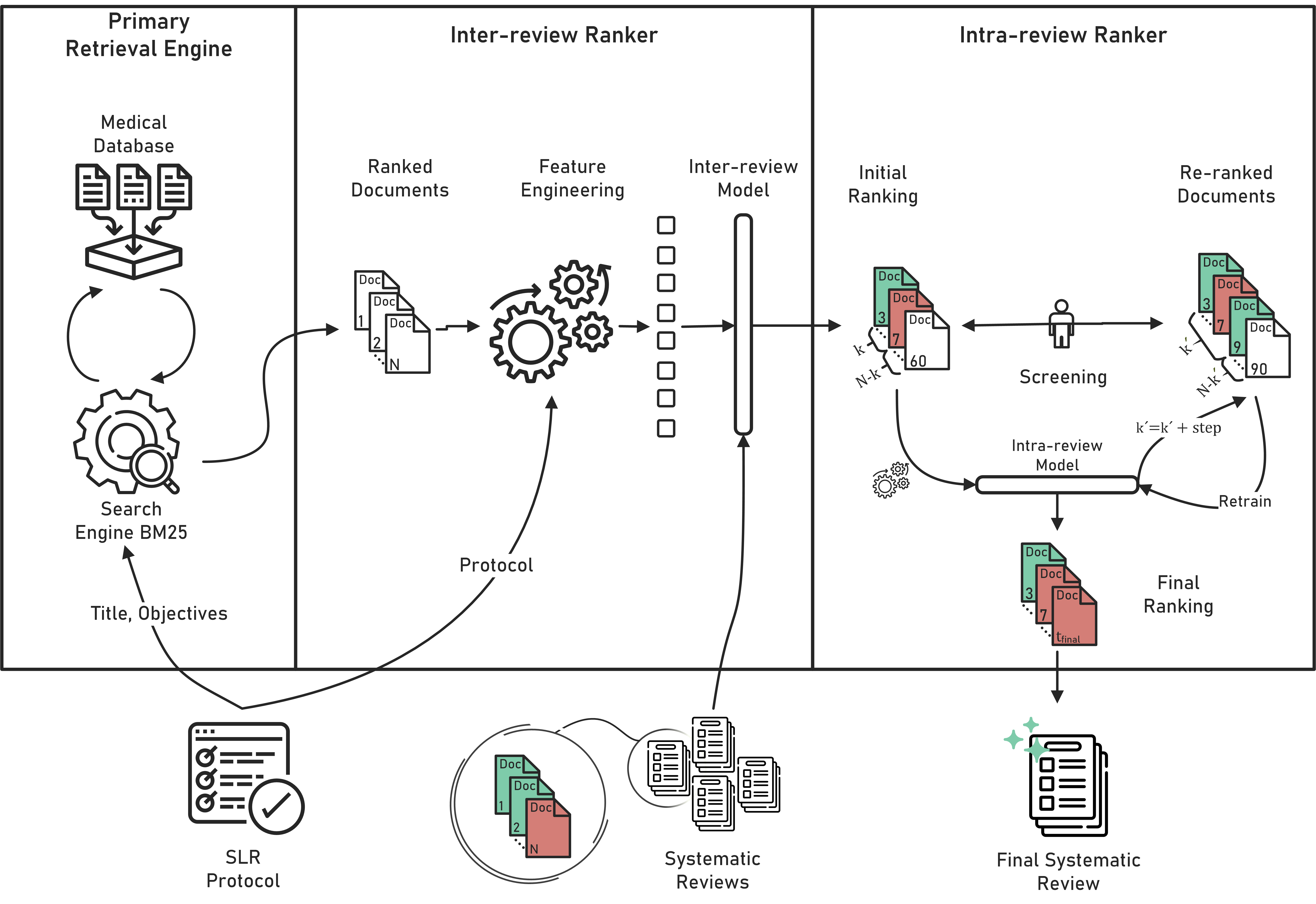}
    \caption{Our hybrid learning pipeline for screening prioritization in systematic literature reviews.}
    \label{fig:methodology}
\end{figure}

This section provides a detailed description of our hybrid learning approach for screening prioritization in systematic reviews. Our approach does not require the construction of a Boolean query by specialists, and consists of three consecutive components: initial retrieval, inter-review ranking, and intra-review ranking. Figure \ref{fig:methodology} illustrates our approach in detail.  

\subsection{Primary Retrieval Engine}
In the first step of our approach, an initial retrieval of the relevant documents is performed using a traditional IR system based on the BM25 score. The title and the objectives of the systematic review, as defined in its protocol, are separately given as queries to the IR system. The final ranked list of documents is the combination of the retrieved documents from both queries. The normalized score for each query are summed to produce the final ranking score for each document. The aim of this component is to retrieve a large number of documents from a database in order to achieve very high recall, while at the same time reducing the number of documents processed in the following steps. In the case of a systematic review, the primary retrieval engine decreases the possibly relevant documents in the biomedical databases from millions to tens of thousands. 

\subsection{Inter-Review Ranker}
The second component of our approach, the inter-review ranker, aims to bring the relevant documents at higher ranking position. This step includes a learning-to-rank (LTR) model that ranks the set of documents retrieved by the primary retrieval engine, according to their relevance and importance for the review topic. Each document is represented as a feature vector, where each feature indicates how relevant the document is with respect to the review topic. The LTR model is trained on previously produced SLRs. The main idea behind this model is that it can grasp the knowledge across different SLRs and then be able to produce an efficient document ranking for an unknown SLR. The list of features includes traditional scoring functions such as BM25 and TF-IDF, LETOR inspired features \citep{qin2010letor} and semantic features using Word2Vec \citep{mikolov2013distributed} and a novel feature deriving from Sent2Vec \citep{pgj2017unsup}. The features are computed using the different fields of the systematic review's protocol, and the title and abstract of the documents. The full list of features is presented in Table \ref{table:inter_review_features}. Details about the features are given below:

\begin{enumerate}
    \item We consider two fields of a document $d$: the title and the abstract. Column ``Document field(s)'' indicates whether these fields are used separately (,) or concatenated into a single string (+).
    \item We consider multiple protocol fields based on the type of the systematic review. We denote a protocol field as $p$. In our study, 4 different types of systematic reviews are considered; Diagnostic Test Accuracy (DTA), Intervention, Qualitative, and Prognosis. All types include a Title, Objectives, Types of Studies, and Types of Participants fields. DTA reviews further include the fields {\em Index Tests}, {\em Target Conditions}, and {\em Reference Standards}. Intervention reviews include the fields {\em Types of Intervention} and {\em Types of Outcome Measures} and Prognosis reviews include the {\em Type of Outcome Measures} field.
    \item The number of occurrences of a protocol field's token $p_i$ in a document is denoted as $c(p_i,d)$.
   \item The BM25 score is computed as in \citep{Robertson2010}.
   \item Singular Value Decomposition (SVD) is performed upon the tf-idf. The cosine similarity is estimated from the reduced vectors of the two fields.
   \item The Word Mover's Distance (WMD) of the word vectors is computed as in \citep{Kusner2015}.
   \item Sentence embeddings are produced by a Sent2Vec pre-trained model \citep{chen2019biosentvec}.

\end{enumerate}

\begin{table}[!ht]
    \label{table:inter_review_features}
    \centering
    \caption{Set of features employed by the inter-review ranker.}

	\renewcommand{\arraystretch}{1.3}
	\resizebox{\textwidth}{!}{%
	\begin{tabular}{cccc}
		\hline
		~ID~ & Description & ~Protocol Field(s) & ~Document field(s)~\\ \hline
		1-18 & BM25 & All & Title, Abstract \\ 
		19-36 & $\log$(BM25) & All & Title, Abstract \\
		37-54 & cos(tf-idf) & All & Title, Abstract \\ 
        55-58 & $\sum_{p_i \in p \cap d} c(p_i,d)$ & Title, Objectives & Title, Abstract \\
        59-62 & $\log\sum_{p_i \in p \cap d} c(p_i,d)$ & Title, Objectives & Title, Abstract \\
        63 & BM25 & Title + Objectives & Title + Abstract \\
        64 & Z-Score(BM25) & Title + Objectives & Title + Abstract \\
        65 & $\frac{|p \cap d|}{|p|}$ & Title + Objectives & Title + Abstract \\
        66 & $\frac{|p^{(b)} \cap  d^{(b)}|}{|p^{(b)}|}$ & Title + Objectives & Title + Abstract \\ 
        67 & $\sum($idf$(p_i \in t \cap d))$ & Title + Objectives & Title + Abstract \\
        68-69  & $cos($SVD$($tf-idf$))$ & Title + Objectives & Title, Abstract \\
        70-71 & WMD$($Word2Vec$)$ & Title, Objectives & Title + Abstract \\
        72 & $cos($Sent2Vec$)$ & Title + Objective & Title + Abstract \\
        \hline
	\end{tabular}%
	}
\end{table}

\subsection{Intra-Review Ranker}
The intra-review ranker is the last component and last step of our approach and employs the screening process conducted by the researcher, during the appraisal stage. This step consists of a simple supervised learning model that is continuously (re)trained based on the reviewer's relevance feedback. For this classifier, sentence embeddings were extracted, from a pre-trained Sent2Vec model, to represent the documents. To the best of our knowledge we are the first consider such embeddings for TAR. Initially, the intra-review ranker is trained on the top-$k$ documents as ranked by the inter-review ranker and assessed by the reviewer. If no relevant documents are found in the top-$k$ documents, the review continues until the training set consists of both relevant and irrelevant documents. Then, iteratively re-ranks the rest of the documents, expanding the training set of the intra-review model with the top-ranked document, until the whole list has been added to the training set or a certain threshold is reached. The expansion of the training set is configured by 4 parameters. Two thresholds, $t_{\textrm{init}}$ and $t_{\textrm{final}}$, are defined. After training with the initial k-documents, the training set is expanded until $t_{\textrm{init}}$ is reached using a step $s_{\textrm{init}}$. Then, $s_{\textrm{init}}$ is increased to $s_{\textrm{final}}$ and the expansion of the training set continues until $t_{\textrm{final}}$. This iterative feedback and re-ranking mechanism is described in detail in Algorithm \ref{alg:rerank_intra}.

The use of different steps and thresholds reduces the cost of feedback and the time needed to produce predictions since the classifier is considered sufficiently trained when the training set has reached a certain number of documents. Moreover, this procedure allows the researcher to set a specific number of documents to be assessed, while taking into consideration the human resources and cost required. 

\begin{algorithm}[!ht]
\SetKwInOut{Input}{Input}
\SetKwInOut{Output}{Output}
\Input{The ranked documents $R$, of length $n$, as produced by the inter-review ranker, initial training step $k$, initial training step $s_{\textrm{init}}$, secondary training step $s_{\textrm{final}}$, step change threshold $t_{\textrm{init}}$, final threshold $t_{\textrm{final}}$ (optional)}
\Output{Final ranking of documents $R$ - $finalRanking$ }
$finalRanking \gets \emptyset $\;

\For{$i=1$ \textbf{to} $k$}{
	$finalRanking_i \gets finalRanking_i \cup R_i $\;
}
$k' \gets k$\;
\While{$finalRanking \textsf{\upshape ~does not contain both \textbf{relevant} and \textbf{irrelevant} documents}$}
{	
	$k' \gets k' + 1$\;
	$finalRanking_{k'}= R_{k'}$\;
}
\While{$|finalRanking| \neq$ $n$ \textbf{\upshape AND} $|finalRanking| \neq t_{final}$}{
    
	$train(finalRanking)$ \tcp*{Train a classifier by asking for relevance for these documents}
    
    $localRanking = rerank(R - finalRanking)$ \tcp*{Rerank the rest of the initial list $R$ based on the probabilities of the classifier}
    \uIf{$|finalRanking| < t_{\textrm{init}}$}
    	{$s = s_{\textrm{init}}$\;}
     \Else
     	{$s = s_{\textrm{final}}$\;}
   	\For{$i=k'$ \textbf{to} $k' + s$}
    {$finalRanking_i \gets localRanking_{i-k'} $\;}
				
	}
    \Return{$finalRanking$\;}

\caption{Iterative relevance feedback algorithm of the intra-review ranker}
\label{alg:rerank_intra}
\end{algorithm}

\section{Empirical Study}
\label{sec:empirical}

This section describes the dataset we used for our study and details the updates we implemented to it. Furthermore, it presents our evaluation process for each of the individual components of our hybrid learning approach and specifies the parameters and tools used for our experiments. Finally, it discusses and compares our results with other approaches presented in the past. 

\subsection{Data}
Our data come from Task 2, TAR in Empirical Medicine, of CLEF e-health lab series from 2019, which extends the data of the 2017 and 2018 versions of the lab. \citep{kanoulas2017clef,Kanoulas2018,kanoulas2019clef}. The training set consists of 90 systematic reviews: 70 Diagnostic Test Accuracy (DTA) studies, and 20 Intervention studies. The test set includes 7 DTA, 16 Intervention, 2 Qualitative, and 1 Prognosis studies. Each SLR in the dataset includes its protocol and two Qrel files (list of relevant documents) for the abstract- and content-level assessment respectively. All the SLRs can be found in the Cochrane Library\footnote{\url{https://www.cochranelibrary.com/}} and the initial dataset that was provided by the organizers of the task is available on GitHub\footnote{\url{https://github.com/CLEF-TAR/tar/tree/master/2019-TAR}}.

After experimenting with the dataset, we noticed several issues both in the format of the dataset (i.e. misspelled tags in XML and folder names) and the integrity of the relative documents (i.e. SLRs are updated over time, have very few relevant documents, SLRs existing in both train and test set, deleted PMIDs and others). Therefore, we've updated the dataset by fixing all the format issues that we found and updated the content-level qrels by scraping the Cochrane library website. The abstract-level qrels could not be updated since documents returned by the boolean query are not available online and were initially provided by the organizers. We aimed to provide an up-to-date version of the dataset that will engage and motivate more researchers on technology assisted reviews.

For updating the qrels we scraped the included studies from the reviews' web page for each of the total 116 SLRs\footnote{Last accessed on April 11th, 2020}. The majority of the referenced studies included the corresponding PMID. For those references missing the PMID we followed a similar procedure as the organizers of the e-health lab \citep{Kanoulas2018}. The title of the reference with the missing PMID was submitted to the PubMed Search Engine\footnote{\url{https://pubmed.ncbi.nlm.nih.gov/}}. If there was a match, the PMID of the returned document was included, if not, the returned result was examined further until a match was found. All other studies, without a match, were discarded under the assumption that these are not contained in PubMed.

The updated qrels contain 14.29\% more documents than the previous qrels with 628 new PMIDs and 126 discarded PMIDs. The updated GitHub repository\footnote{\url{https://github.com/sakrifor/tar}} contains the updated qrel files for the 2019 dataset, several format corrections, statistics on the new dataset and the references scraped from the Cochrane Library website along with their PubMed link and PMID.

\subsection{Hybrid Learning Implementation}
\label{sec:method_settings}
In this section, we describe all the different tools, libraries, and settings we used in order to implement and evaluate the approach presented in Section \ref{sec:methodology}. 

As our primary retrieval engine, we use the Elasticsearch search engine\footnote{\url{https://www.elastic.co/elasticsearch/}} along with the PubMed 2019 annual baseline\footnote{\url{ftp://ftp.ncbi.nlm.nih.gov/pubmed/baseline/}}. Our index includes more than $29.100.000$ articles. For each systematic review, we submit two queries to the Elasticsearch using the title or the objectives of an SLR's protocol. The queries are formulated using a dictionary created from the most common words in PubMed. Words with frequencies higher than 50\% or with less than 10 counts were discarded. The term threshold was set to 100k. The returned documents from both queries are combined into a single list scored by the normalized rank of the individual lists. For each SLR, 100k documents were returned with the intention to achieve high total recall. The final query formulation (i.e. combination of two queries) was selected based on recall on the train set, after experimenting with other queries, such as single query, concatenation of queries, and different dictionaries. 

For the second component of our approach, the inter-review ranker, we use the LambdaMart algorithm as implemented in the XGBoost (XGBRanker, default parameters, tree\_method=``gpu\_hist")~\citep{chen2016xgboost}. For each document, retrieved by the primary retrieval engine, a vector with the features from Table \ref{table:inter_review_features} is created. The TF-IDF was constructed using the same vocabulary as in the initial ranking. The pre-trained Word2Vec model from the BioASQ challenge, trained on the PubMed abstract \citep{pavlopoulos2014continuous}, and the BioSentVec model~\citep{chen2019biosentvec}, trained both on PubMed and MIMIC III clinical notes, were adopted for the Word2Vec and Sent2Vec model respectively.

Finally, we employ the intra-review ranker. The classifier used is a Linear SVM (scikit-learn library, regularization parameter $C=0.5$) and the features for each document were extracted using the pre-trained BioSent2Vec model \citep{chen2019biosentvec} applied on the concatenation of the title and the abstract of the article. As default parameters, we set $k=10,s_{\textrm{init}}=1,t_{\textrm{init}}=200,s_{\textrm{final}}=50$ and $t_{\textrm{final}}=1000$. Thus, the top-10 ranked documents are obtained from the inter-review ranker and screened by the reviewer. The intra-review ranker is then trained and the document ranking is updated. The reviewer continues to screen the documents and for each document screened a new ranking is produced by continuously retraining the intra-review model. The process continues until 200 documents are reviewed. After that, the intra-review model retraining and update of the ranking of the documents occur every 50 documents reviewed. The process is repeated until the final threshold of 1000 documents is reached. We assume that reviewers always classify correctly a document as relevant or not. 

\subsection{Evaluation and Results}

Our evaluation process is two-fold. We first evaluate the different components of our hybrid learning approach, presented in Section \ref{sec:methodology}, and then, we compare our results with other state-of-the-art approaches. We aim to determine whether the retrieval of relevant documents for technology assisted reviews is pragmatic and efficient without the use of a boolean query. In both cases, we use the updated content-level relevant documents, final studies included in the studies after the full-text screening, and we evaluate our approach using both the train and the test set.

\subsubsection{Components Validation}

Table \ref{tab:method_results} presents the results for each of the components of our approach using the train set. We used the leave-one-out method on the 90 SLRs included in the training set and evaluation measures are computed using the script provided by the organizers of the Task 2 of CLEF e-health lab \citep{kanoulas2017clef}. The table includes the Recall@threshold where threshold$=\{10,50,100,5000\}$ documents, Mean Average Precision (MAP), Work Saved over Sampling at 100\% recall (WSS@100) \citep{cohen2006reducing} and last\_rel. The metric last\_rel is the minimum number of documents returned to retrieve all relevant documents. The Recall@100k, which is the maximum recall our approach can achieve, is 0.9224. We consider recall@5000 documents as the maximum valid threshold. We believe a higher threshold than 5000 to be unrealistic in a real-world case scenario.   


\begin{table}[]
\centering
\caption{Comparison of the three different components of our Hybrid learning approach using the training set.}
\label{tab:method_results}
\renewcommand{\arraystretch}{1.3}
\resizebox{\textwidth}{!}{%
\begin{tabular}{lcccccccc}
\toprule
 & Recall@10 & Recall@50 & Recall@100 & Recall@5000 & MAP & WSS@100 & last\_rel \\ \midrule
Initial Retrieval &  0.0297 & 0.1213 & 0.1866 & 0.7401 & 0.0565 & 0.1736 & 33258 \\
Inter-Review &  0.0421  & 0.1356 & 0.209 & 0.8239 & 0.0710 & 0.5917 & 16693 \\
Intra-Review &  0.0421 & 0.1765 & 0.3411 & 0.9100 & 0.1120 & 0.6137 & 2391 \\
\bottomrule
\end{tabular}%
}
\end{table}

In total, $4.645.817$ articles were retrieved, by the primary retrieval engine (100k documents per SLR), after duplicates removal. Comparing the primary retrieval engine, with the inter-review ranker, we first notice that retrieving such an amount of documents helps us reach very high recall which is our main concern for this problem. However, the primary retrieval engine the Recall@5000 remains low (0.7401). The gap between those thresholds is diminished by the next step of our approach, the inter-review ranker, which reaches a recall@5000 of 0.8339. Till this step, no relevance feedback from the reviewers is used. A similar increase occurs also at thresholds 10, 50, and 100 of recall. Likewise, the MAP, WSS@100, and last\_rel metrics are also improved. Notably, the last\_rel is decreased by more than 16000 documents. This is a clear indication of the significance of the inter-review ranking on bringing all the relevant documents higher in the ranking and, thus, more accessible by the reviewer on the final stage of our approach.

Regarding the intra-review ranking, we observe an additional improvement in all metrics. A great indication of this improvement is the last\_rel metric, where the minimum number of documents returned to retrieve all relevant documents is decreased to 2391 from 16693 documents, and the recall@5000 reaches a very high recall, 0.9100, which is suited for this task.

\begin{table}[]
\centering
\caption{Comparison of the three different components of our Hybrid Learning approach using the test set.}
\label{tab:test_method_results}
\renewcommand{\arraystretch}{1.3}
\resizebox{\textwidth}{!}{%
\begin{tabular}{lccccccc}
\toprule
 & Recall@10 & Recall@50 & Recall@100 & Recall@5000 & MAP & WSS@100 & last\_rel \\ \midrule
Initial Retrieval & 0.012 & 0.0552 & 0.0898 & 0.5057 & 0.0405 & 0.219 & 50218 \\
Inter-Review & 0.0113 & 0.0566 & 0.0976 & 0.5573 & 0.049 & 0.2788 & 18325 \\
Intra-Review & 0.0113 & 0.0827 & 0.1938 & 0.7235 & 0.0985 & 0.2976 & 5638 \\
\bottomrule
\end{tabular}%
}
\end{table}

We also test our approach on the dataset's test set. Table \ref{tab:test_method_results} presents the results of each of the components of our hybrid learning approach. The inter-review ranker was trained with the 90 SLRs of the training set and the parameters were set as described in Section \ref{sec:method_settings}. The initial retrieval component achieves a recall@100k documents of 0.7595, and a total of 2.172.204 were retrieved. We first notice that the recall@100k is much lower compared to the results of the training set. However, all the other metrics continue to improve as our hybrid methodology progresses.

\begin{table}[]
\centering
\caption{Comparing DTA, Intervention, Prognosis and Qualitative SLRs on the three different component of our Hybrid Learning approach.}
\label{tab:method_types}
\renewcommand{\arraystretch}{1.3}
\resizebox{\textwidth}{!}{%
\begin{tabular}{lcccccc}
\toprule
 & \multicolumn{2}{c}{Primary Retrieval Engine} & \multicolumn{2}{c}{Inter-Review} & \multicolumn{2}{c}{Intra-Review} \\
 & Recall@5000 & MAP & Recall@5000 & MAP & Recall@5000 & MAP \\ \midrule
DTA & 0.6893 & 0.0539 & 0.8380 & 0.0702 & 0.9497 & 0.1420 \\
Intervention & 0.6403 & 0.0417 & 0.6460 & 0.0480 & 0.8230 & 0.0923 \\
Prognosis & 0.0312 & 0.0004 & 0.0623 & 0.0110 & 0.2648 & 0.0208 \\
Qualitative & 0.4500 & 0.0033 & 0.4835 & 0.0074 & 0.7500 & 0.0342 \\
\bottomrule
\end{tabular}%
}
\end{table}

We look further into this issue of the low recall by reporting the results separately for each type of SLR in Table \ref{tab:method_types}. The Recall@100k for DTA, Intervention, Prognosis, and Qualitative is 0.9385, 0.8797, 0.2897, and 0.7667 respectively. We notice that Prognosis achieves much lower results on all the metrics compared to the other types, due to very low recall from the initial retrieval stage. Our dataset consists only of 1 prognosis SLR with 321 relevant articles out of a total of 1414 documents across all types of SLRs which justifies the low metric scores when averaged across all SLRs. Further research is needed to identify any possible oddities on the nature of prognosis SLRs. DTA SLRs achieve the highest results in all cases that can be attributed to the fact that the train set consists mostly (70/90) of DTA SLRs. Our assumption that the types of SLRs in the training set greatly affects the inter-review ranker's performance is also supported by the much lower scores of the Qualitative SLRs.

\subsubsection{Feature importance \& Parameter setting}
\label{sec:feat_n_param}

In an attempt to better understand how our ranker components make predictions we look into the top-15 features of the inter-review ranker and how the predictions are affected by the different threshold parameters of the intra-review ranker.

Table \ref{tab:feature_importance} presents the top-15 features sorted by the feature's importance scores from the XGBoost ranker used in the inter-review model. We first notice that traditional features, such as TF-IDF and BM25 based features, populate most of the positions in the list, with the highest score being held by the Z-score of BM25. The LETOR inspired features hold a lower position in our ranking while the newly introduced feature, the cosine similarity of Sent2Vec embeddings, is ranked second. This upholds to the trend of the significance of semantic word/sentence embeddings. Finally, the field used either in document or protocol level doesn't imply a notable influence on the final performance since all fields appear in different positions in the list.

\begin{table}[]
\caption{Top-15 features as scored by the XGBoost in the inter-review ranker.}
\label{tab:feature_importance}
\begin{tabular}{ccccc}
\toprule
Rank & Feature & Protocol Field(s) & Document Field(s) & Score \\ \midrule
1 & Z-Score(BM25) & Title + Objectives & Title + Abstract & 57.0533 \\
2 & cos(Sent2Vec) & Title + Objectives & Title + Abstract & 5.4320 \\
3 & cos(tf-idf) & Types of Participants & Title & 2.2487 \\
4 & $\log$(BM25) & Objectives & Abstract & 1.8549 \\
5 & BM25 & Title & Title & 1.4538 \\
6 & cos(tf-idf) & Title & Title & 1.4533 \\
7 & cos(tf-idf) & Type of Studies & Title & 1.3359 \\
8 & cos(tf-idf) & Objectives & Abstract & 1.3172 \\
9 & BM25 & Title & Abstract & 1.0596 \\
10 & $\log$(BM25) & Title & Abstract & 1.0519 \\
11 & $\log$(BM25) & Title & Title & 0.9912 \\
12 & WMD(Word2Vec) & Objectives & Title + Abstract & 0.9484 \\
13 & $cos($SVD$($tf-idf$))$ & Title + Objectives & Abstract & 0.9255 \\
14 & $\log\sum_{p_i \in p \cap d} c(p_i,d)$ & Objectives & Abstract & 0.8032 \\
15 & $\frac{|p \cap d|}{|p|}$ & Title + Objectives & Title + Abstract & 0.6795 \\ \bottomrule
\end{tabular}
\end{table}

Table \ref{tab:parameter_setting} shows the results of the intra-review ranking on the training set using different thresholds, $t_{\textrm{init}}, t_{\textrm{final}}$. We keep the rest of the parameters constant with $k=10$, $s_{\textrm{init}}=1$, and $s_{\textrm{final}}=50$, since $k$ is increased if both relevant and irrelevant documents are not present and small changes in $s_{\textrm{init}}, s_{\textrm{final}}$ do not affect the results. Both step parameters are associated with the computational cost of this methodology and their effect is out of the scope of this paper. In an aim to keep the computational cost in coordination with a real case scenario, we adjust the parameters to the above values so as to have continuous and constant retraining, until a relatively low threshold is reached, and more ``distant in time" re-trainings after our training is considered to be enough for effective ranking. After taking into consideration the above, we conclude that the best parameters are $t_{\textrm{init}} = 200, t_{\textrm{final}} = 2000$ since they achieve the second-best performance in all metrics and the relevance feedback needed is much lower than this of the best performing parameters (2000 documents vs 5000 documents). In a real-world case, all the parameters can be adjusted according to the resources and the outcome during the assessment.  

\begin{table}[]
\centering
\caption{The intra-review ranker using different thresholds ($t_{\textrm{init}}, t_{\textrm{final}}$ while keeping the step parameters fixed ($k=10, s_{\textrm{init}}=1,s_{\textrm{final}}=50$)}
\label{tab:parameter_setting}
\renewcommand{\arraystretch}{1.3}
\begin{tabular}{llcccc}
\toprule
$t_{\textrm{init}}$ & $t_{\textrm{final}}$  & Recall@5000 & MAP & WSS@100 & last\_rel \\ \midrule
200 & 500  & 0.8965 & 0.1142 & 0.5927 & 3671 \\
200 & 1000 & 0.9100 & 0.1120 & 0.6137 & 2391 \\
200 & 2000 & 0.9127* & 0.1148* & 0.6003* & 2274* \\
200 & 5000 & \textbf{0.9135} & 0.1148* & \textbf{0.6004} & \textbf{2098}  \\
300 & 1000 & 0.9046 & 0.1155 & 0.5997 & 2599 \\
500 & 1000 & 0.9042 & \textbf{0.1157} & 0.5998 & 2616 \\
\bottomrule
\end{tabular}
\end{table}

\subsubsection{Comparison with state-of-the-art method and a baseline}

We compare our approach with three other approaches, the AUTO-TAR BMI method by \cite{Cormack2018}, which is considered state-of-the-art, a variation of our Hybrid methodology (Hybrid TFIDF) which we have previously submitted at Task 2 of CLEF e-health lab \citep{minas2018aristotle} and a simple PubMed retrieval method. 

The AUTO-TAR BMI method uses a continuous active learning method where random documents are selected and evaluated by the reviewer. A constantly increasing number of documents are added to the ranking and a new training set is continuously created. The process ends when all the documents have been screened. For the implementation of the AUTO-TAR BMI, we followed Algorithm 1 as described in \cite{Cormack2018}. The Hybrid variation method is similar to the one presented in \cite{lagopoulos2018learning} which uses TF-IDF instead of the Sent2Vec for the representation of the documents in the intra-review model. Finally, for the PubMed method, we query the PubMed database\footnote{\url{https://pubmed.ncbi.nlm.nih.gov/}}, using the Entrez Programming Utilities, with the title and the objective of each SLR. The normalized position of both queries is combined to a single ranking. For the current Hybrid approach (Hybrid Sent2Vec) we set the same parameters as in Section \ref{sec:method_settings} but we choose the best thresholds for intra-review ranker as computed in Section \ref{sec:feat_n_param} with $t_{\textrm{init}} = 200, t_{\textrm{final}} = 2000$.

Tables \ref{tab:all_methods_train} and \ref{tab:all_methods_test} present the results for each of the methods using the train set (leave-one-out) and the test set, respectively. Initially, we notice that our approach outperforms all the other approaches in almost all metrics, with the AUTO-TAR approach following. Specifically, in the training set, our hybrid approach greatly outperforms AUTO-TAR in all the recall thresholds, the MAP, and the last\_rel metrics, while AUTO-TAR marginally surpasses our method at WSS@100. Our approach also outperforms AUTO-TAR in the test set; nevertheless, the difference is marginal in almost all metrics with recall at thresholds 50, 100, and 5000 showing significant difference. However, hybrid learning uses much fewer resources in terms of relevance feedback compared to AUTO-TAR
which asks for assessment for all 5000 documents. The addition of a stopping method or criterion could possibly distinguish the performance of the two approaches further and will be considered in future work.

\begin{table}[]
\centering
\caption{Comparing Hybrid Learning with other approaches using the training set of 90 SLRs in a Leave-One-Out fashion.}
\label{tab:all_methods_train}
\renewcommand{\arraystretch}{1.3}
\resizebox{\textwidth}{!}{%
\begin{tabular}{lccccccc}
\toprule
 & Recall@10 & Recall@50 & Recall@100 & Recall@5000 & MAP & WSS@100 & last\_rel \\ \midrule
PubMed & 0.0147 & 0.0502 & 0.0765 & 0.4442 & 0.0207 & 0.1639 & 12910 \\
Auto-TAR & 0.0031 & 0.0336 & 0.0846 & 0.8849 & 0.043 &\textbf{ 0.6181} & 3000 \\
Hybrid (TFIDF) & 0.0421 & 0.0537 & 0.0595 & 0.5620 & 0.0298 & 0.3627 & 5830  \\ 
Hybrid (Sent2Vec) &\textbf{ 0.0421} & \textbf{0.1761} & \textbf{0.3272} &\textbf{ 0.9127 }& \textbf{0.1148} & 0.6003 & \textbf{2274} \\

\bottomrule
\end{tabular}%
}
\end{table}

\begin{table}[]
\centering
\caption{Comparing Hybrid Learning with other approaches on the test set of 26 SLRs.}
\label{tab:all_methods_test}
\renewcommand{\arraystretch}{1.3}
\resizebox{\textwidth}{!}{%
\begin{tabular}{lccccccc}
\toprule
 & Recall@10 & Recall@50 & Recall@100 & Recall@5000 & MAP & WSS@100 & last\_rel \\ \midrule
PubMed & 0.0064 & 0.0347 & 0.0495 & 0.3140	& 0.0155 & 0.0719 & 8658 \\
AUTO-TAR & \textbf{0.0149} & 0.0622 & 0.1047 & 0.7037 & 0.0933 & 0.3010 & 4018 \\
Hybrid (TFIDF) & 0.0113 & 0.0410 & 0.0830 & 0.4860 & 0.0725 & 0.1753 & 6870 \\
Hybrid (Sent2Vec) & 0.0113 & \textbf{0.0849} & \textbf{0.1881} & \textbf{0.7327} & \textbf{0.0970} & \textbf{0.3054} & \textbf{3976} \\   \bottomrule
\end{tabular}%
}
\end{table}

\begin{figure}[ht]
    \centering
    \includegraphics[width=0.8\textwidth]{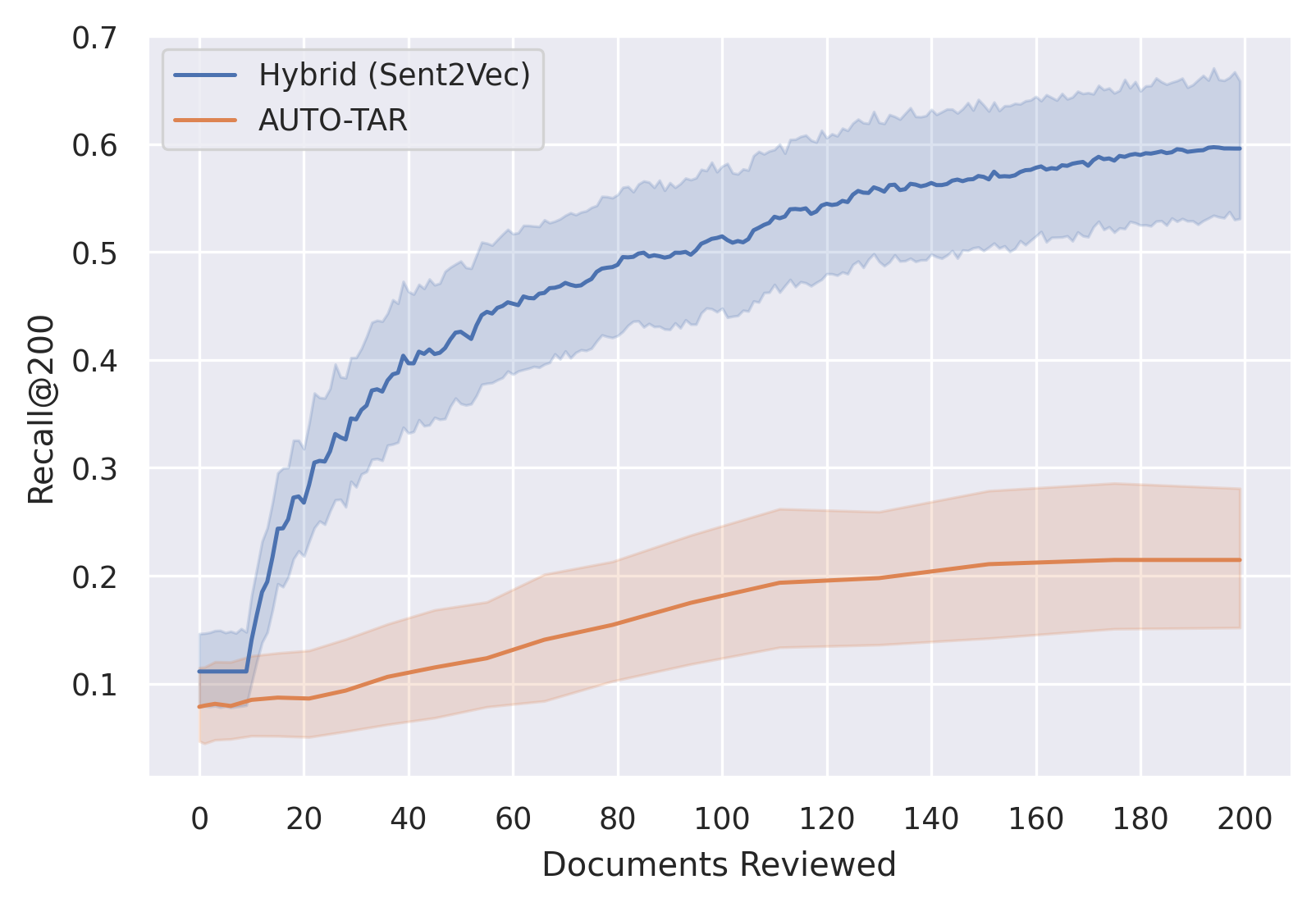}
    \caption{Graph showing macro recall@200 for the Hybrid learning and AUTO-TAR approaches along with the 95\% confidence interval at the train set.}
    \label{fig:train_macro}
\end{figure}

\begin{figure}[ht]
    \centering
    \includegraphics[width=0.8\textwidth]{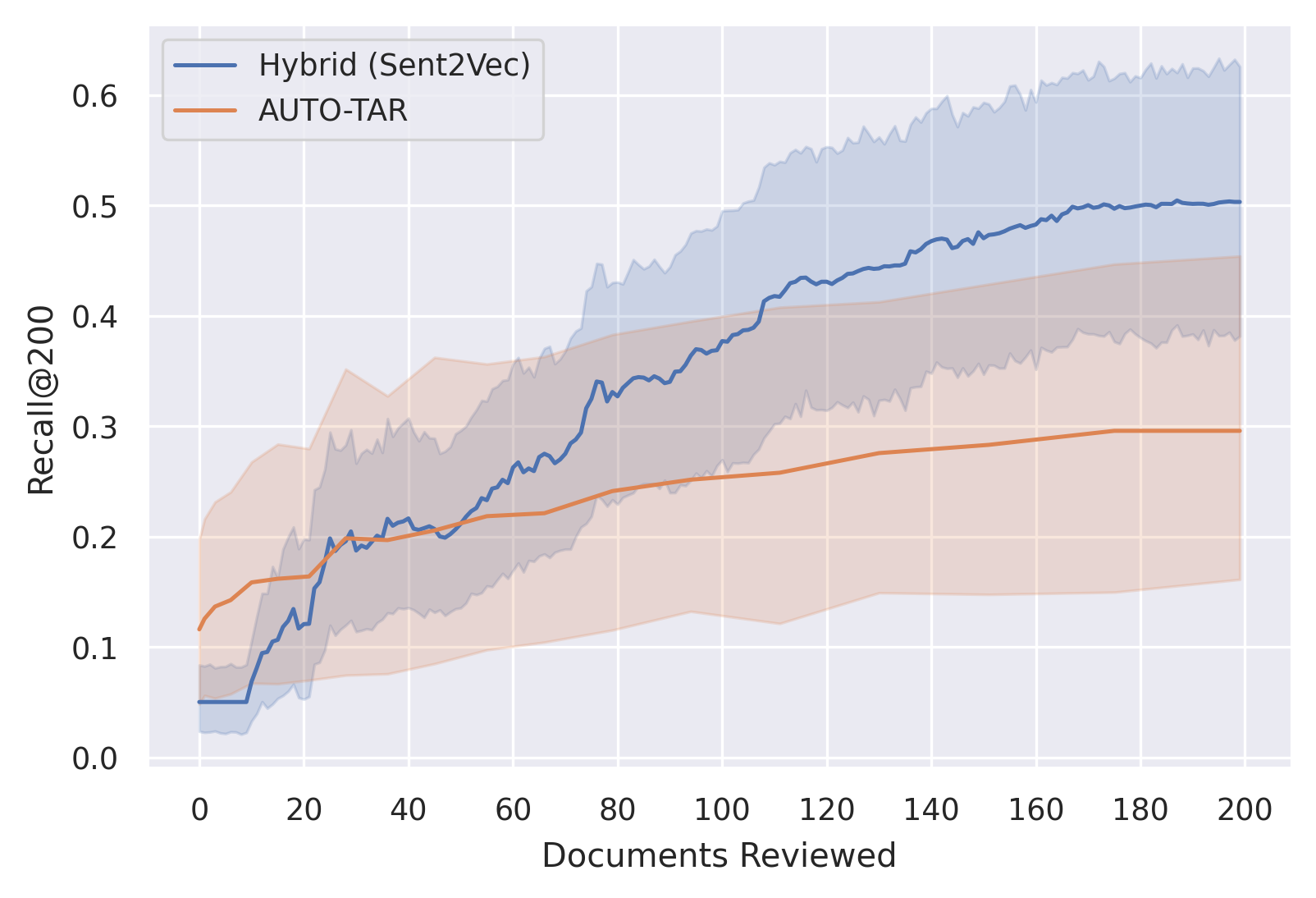}
    \caption{Graph showing macro recall@200 for the Hybrid learning and AUTO-TAR approaches along with the 95\% confidence interval at the test set.}
    \label{fig:test_macro}
\end{figure}

We further examine how our approach compares to others by measuring recall@200 during the review process. Figures \ref{fig:train_macro},\ref{fig:test_macro} show macro recall@200 for the AUTO-TAR and hybrid approach along with the 95\% confidence interval for the train set and test set, respectively. We first spot large fluctuations in the hybrid approach compared to AUTO-TAR. This is due to the re-training process that the AUTO-TAR follows, which doesn't re-ranks the documents after each document has been reviewed, as the hybrid learning, but the retraining occurs after $B$ documents have been reviewed, where $B$ constantly increases with a rate of $\ceil*{\frac{B}{10}}$. Moreover, we notice that hybrid learning increases rapidly after the first few documents have been reviewed and it reaches high recall when 200 have been reviewed compared to AUTO-TAR which maintains a low increase rate. Both observations indicate that the intra-review ranking exploits better the relevance feedback compared to AUTO-TAR and is better suited to a live system, where relevant documents will appear sooner to the reviewer and the relevance feedback will have immediate effects.

\section{Conclusion \& Feature Work}
\label{sec:conclusion}
In this work, we introduced a novel approach to screening prioritization for systematic reviews which consists of three consecutive components. Our approach doesn't make use of a boolean query and solely relies on the protocol of an SLR, assisting researchers in the preparation, retrieval, and appraisal stages of preparing an SLR. Furthermore, our inter-review ranking component learns from other reviews which enables it to adapt to other types of reviews and possibly achieve better performance. Additionally, hybrid learning incorporates novel elements, such as sentence embeddings, which render our approach as a strong and present-day baseline. 

We performed an empirical study on an updated version of the dataset provided by Task II of CLEF e-health lab which we also make publicly available. Our experiments show that our hybrid approach outperforms the state-of-the-art and that is suitable for a real-world case scenario. Finally, our study uncovers a simple, transparent, and effective baseline for screening prioritization.

In the future, we plan to further investigate how we can improve our inter-review component and minimize the relevance feedback needed to achieve total-recall and high precision by using active learning techniques and zero-shot learning. Moreover, we will look into stopping methods and how they affect performance. As a final step, we intend to build and test an application that employs our approach and benefits researchers starting a systematic review.

\bibliographystyle{spbasic}      
\bibliography{mybibfile}   

\begin{thebibliography}{57}
\providecommand{\natexlab}[1]{#1}
\providecommand{\url}[1]{{#1}}
\providecommand{\urlprefix}{URL }
\expandafter\ifx\csname urlstyle\endcsname\relax
  \providecommand{\doi}[1]{DOI~\discretionary{}{}{}#1}\else
  \providecommand{\doi}{DOI~\discretionary{}{}{}\begingroup
  \urlstyle{rm}\Url}\fi
\providecommand{\eprint}[2][]{\url{#2}}

\bibitem[{Alharbi and Stevenson(2017)}]{Alharbi2017}
Alharbi A, Stevenson M (2017) {Ranking abstracts to identify relevant evidence
  for systematic reviews: The University of Sheffield's approach to CLEF
  eHealth 2017 Task 2: Working notes for CLEF 2017}. In: CEUR Workshop
  Proceedings, vol 1866

\bibitem[{Alharbi and Stevenson(2019)}]{Alharbi2019}
Alharbi A, Stevenson M (2019) {Ranking studies for systematic reviews using
  query adaptation: University of Sheffield's approach to CLEF eHealth 2019
  task 2 working notes for CLEF 2019}. In: CEUR Workshop Proceedings, vol 2380

\bibitem[{Alharbi et~al.(2018)Alharbi, Briggs, and Stevenson}]{Alharbi2018}
Alharbi A, Briggs W, Stevenson M (2018) {Retrieving and ranking studies for
  systematic reviews: University of Sheffield's approach to CLEF eHealth 2018
  Task 2}. In: CEUR Workshop Proceedings, vol 2125

\bibitem[{Anagnostou et~al.(2017)Anagnostou, Lagopoulos, Tsoumakas, and
  Vlachavas}]{anagnostou2017combining}
Anagnostou A, Lagopoulos A, Tsoumakas G, Vlachavas I (2017) Combining
  inter-review learning-to-rank and intra-review incremental training for title
  and abstract screening in systematic reviews. In: CEUR Workshop Proceedings,
  Aristotle University of Thessaloniki, vol 1866

\bibitem[{Bekhuis and Demner-Fushman(2010)}]{bekhuis2010towards}
Bekhuis T, Demner-Fushman D (2010) Towards automating the initial screening
  phase of a systematic review. In: MedInfo, pp 146--150

\bibitem[{Bekhuis and Demner-Fushman(2012)}]{bekhuis2012screening}
Bekhuis T, Demner-Fushman D (2012) Screening nonrandomized studies for medical
  systematic reviews: a comparative study of classifiers. Artificial
  intelligence in medicine 55(3):197--207

\bibitem[{Borah et~al.(2017)Borah, Brown, Capers, and
  Kaiser}]{borah2017analysis}
Borah R, Brown AW, Capers PL, Kaiser KA (2017) Analysis of the time and workers
  needed to conduct systematic reviews of medical interventions using data from
  the prospero registry. BMJ open 7(2):e012545

\bibitem[{Chen et~al.(2017)Chen, Chen, Song, Liu, Wang, Hu, He, and
  Yang}]{Chen2017}
Chen J, Chen S, Song Y, Liu H, Wang Y, Hu Q, He L, Yang Y (2017) {ECNU at 2017
  eHealth task 2: Technologically assisted reviews in empirical medicine}. In:
  CEUR Workshop Proceedings, vol 1866

\bibitem[{Chen et~al.(2019)Chen, Peng, and Lu}]{chen2019biosentvec}
Chen Q, Peng Y, Lu Z (2019) Biosentvec: creating sentence embeddings for
  biomedical texts. In: 2019 IEEE International Conference on Healthcare
  Informatics (ICHI), IEEE, pp 1--5

\bibitem[{Chen and Guestrin(2016)}]{chen2016xgboost}
Chen T, Guestrin C (2016) Xgboost: A scalable tree boosting system. In:
  Proceedings of the 22nd acm sigkdd international conference on knowledge
  discovery and data mining, pp 785--794

\bibitem[{Cohen(2006)}]{cohen2006effective}
Cohen AM (2006) An effective general purpose approach for automated biomedical
  document classification. In: AMIA annual symposium proceedings, American
  Medical Informatics Association, vol 2006, p 161

\bibitem[{Cohen and Smalheiser(2018)}]{Cohen2018}
Cohen AM, Smalheiser NR (2018) {UIC/OHSU CLEF 2018 Task 2 diagnostic test
  accuracy ranking using publication type cluster similarity measures}. In:
  CEUR Workshop Proceedings, vol 2125

\bibitem[{Cohen et~al.(2006)Cohen, Hersh, Peterson, and
  Yen}]{cohen2006reducing}
Cohen AM, Hersh WR, Peterson K, Yen PY (2006) Reducing workload in systematic
  review preparation using automated citation classification. Journal of the
  American Medical Informatics Association 13(2):206--219

\bibitem[{Cohen et~al.(2009)Cohen, Ambert, and McDonagh}]{cohen2009cross}
Cohen AM, Ambert K, McDonagh M (2009) Cross-topic learning for work
  prioritization in systematic review creation and update. Journal of the
  American Medical Informatics Association 16(5):690--704

\bibitem[{Cormack and Grossman(2017)}]{Cormack2017}
Cormack GV, Grossman MR (2017) {Technology-assisted review in empirical
  medicine: Waterloo participation in CLEF eHealth 2017}. In: CEUR Workshop
  Proceedings, vol 1866

\bibitem[{Cormack and Grossman(2018)}]{Cormack2018}
Cormack GV, Grossman MR (2018) {Technology-assisted review in empirical
  medicine: Waterloo participation in CLEF eHealth 2018}. CEUR Workshop
  Proceedings 2125

\bibitem[{{Di Nunzio} et~al.(2017){Di Nunzio}, Beghini, Vezzani, and
  Henrot}]{DiNunzio2017}
{Di Nunzio} GM, Beghini F, Vezzani F, Henrot G (2017) {An interactive
  two-dimensional approach to query aspects rewriting in systematic reviews.
  IMS unipd at CLEF eHealth task 2}. CEUR Workshop Proceedings 1866

\bibitem[{{Di Nunzio} et~al.(2018){Di Nunzio}, Ciuffreda, and
  Vezzani}]{DiNunzio2018}
{Di Nunzio} GM, Ciuffreda G, Vezzani F (2018) {Interactive sampling for
  systematic reviews. IMS Unipd at CLEF 2018 eHealth Task 2}. CEUR Workshop
  Proceedings 2125

\bibitem[{Felizardo et~al.(2013)Felizardo, Souza, and
  Maldonado}]{felizardo2013use}
Felizardo KR, Souza SR, Maldonado JC (2013) The use of visual text mining to
  support the study selection activity in systematic literature reviews: a
  replication study. In: 2013 3rd International Workshop on Replication in
  Empirical Software Engineering Research, IEEE, pp 91--100

\bibitem[{Frunza et~al.(2010)Frunza, Inkpen, and Matwin}]{frunza2010building}
Frunza O, Inkpen D, Matwin S (2010) Building systematic reviews using automatic
  text classification techniques. In: Proceedings of the 23rd International
  Conference on Computational Linguistics: Posters, Association for
  Computational Linguistics, pp 303--311

\bibitem[{Frunza et~al.(2011)Frunza, Inkpen, Matwin, Klement, and
  O’blenis}]{frunza2011exploiting}
Frunza O, Inkpen D, Matwin S, Klement W, O’blenis P (2011) Exploiting the
  systematic review protocol for classification of medical abstracts.
  Artificial intelligence in medicine 51(1):17--25

\bibitem[{Higgins et~al.(2019)Higgins, Thomas, Chandler, Cumpston, Li, Page,
  and Welch}]{higgins2019cochrane}
Higgins JP, Thomas J, Chandler J, Cumpston M, Li T, Page MJ, Welch VA (2019)
  Cochrane Handbook for Systematic Reviews of Interventions version 6.0.
  Cochrane, \url{www.training.cochrane.org/handbook}

\bibitem[{Hollmann and Eickhoff(2017)}]{Hollmann2017}
Hollmann N, Eickhoff C (2017) {Ranking and feedback-based stopping for
  recall-centric document retrieval}. In: CEUR Workshop Proceedings, vol 1866

\bibitem[{Kalphov et~al.(2017)Kalphov, Georgiadis, and Azzopardi}]{Kalphov2017}
Kalphov V, Georgiadis G, Azzopardi L (2017) {SiS at CLEF 2017 eHealth tar
  task}. In: CEUR Workshop Proceedings, vol 1866

\bibitem[{Kanoulas et~al.(2017)Kanoulas, Li, Azzopardi, and
  Spijker}]{kanoulas2017clef}
Kanoulas E, Li D, Azzopardi L, Spijker R (2017) Clef 2017 technologically
  assisted reviews in empirical medicine overview. In: CEUR Workshop
  Proceedings, vol 1866, pp 1--29

\bibitem[{Kanoulas et~al.(2018)Kanoulas, Li, Azzopardi, and
  Spijker}]{Kanoulas2018}
Kanoulas E, Li D, Azzopardi L, Spijker R (2018) {CLEF 2018 technologically
  assisted reviews in empirical medicine overview}. In: CEUR Workshop
  Proceedings, vol 2125,
  \urlprefix\url{http://ceur-ws.org/Vol-2125/invited{\_}paper{\_}6.pdf}

\bibitem[{Kanoulas et~al.(2019)Kanoulas, Li, Azzopardi, and
  Spijker}]{kanoulas2019clef}
Kanoulas E, Li D, Azzopardi L, Spijker R (2019) Clef 2019 technology assisted
  reviews in empirical medicine overview. In: CEUR Workshop Proceedings, vol
  2380

\bibitem[{Karimi et~al.(2010)Karimi, Pohl, Scholer, Cavedon, and
  Zobel}]{karimi2010boolean}
Karimi S, Pohl S, Scholer F, Cavedon L, Zobel J (2010) Boolean versus ranked
  querying for biomedical systematic reviews. BMC medical informatics and
  decision making 10(1):58

\bibitem[{Kusner et~al.(2015)Kusner, Sun, Kolkin, and Weinberger}]{Kusner2015}
Kusner MJ, Sun Y, Kolkin NI, Weinberger KQ (2015) {From Word Embeddings To
  Document Distances}. Proceedings of The 32nd International Conference on
  Machine Learning 37:957--966

\bibitem[{Lagopoulos et~al.(2018)Lagopoulos, Anagnostou, Minas, and
  Tsoumakas}]{lagopoulos2018learning}
Lagopoulos A, Anagnostou A, Minas A, Tsoumakas G (2018) Learning-to-rank and
  relevance feedback for literature appraisal in empirical medicine. In:
  International Conference of the Cross-Language Evaluation Forum for European
  Languages, Springer, pp 52--63

\bibitem[{Lee(2017)}]{Lee2017}
Lee GE (2017) {A study of convolutional neural networks for clinical document
  classification in systematic reviews: Sysreview at CLEF eHealth 2017}. In:
  CEUR Workshop Proceedings, vol 1866

\bibitem[{Li and Kanoulas(2019)}]{Li2019}
Li D, Kanoulas E (2019) {Automatic thresholding by sampling documents and
  estimating recall ILPs@UVA at Tar task 2.2}. In: CEUR Workshop Proceedings,
  vol 2380

\bibitem[{Matwin et~al.(2010)Matwin, Kouznetsov, Inkpen, Frunza, and
  O'Blenis}]{matwin2010new}
Matwin S, Kouznetsov A, Inkpen D, Frunza O, O'Blenis P (2010) A new algorithm
  for reducing the workload of experts in performing systematic reviews.
  Journal of the American Medical Informatics Association 17(4):446--453

\bibitem[{Michelson and Reuter(2019)}]{Michelson2019}
Michelson M, Reuter K (2019) {The significant cost of systematic reviews and
  meta-analyses: A call for greater involvement of machine learning to assess
  the promise of clinical trials}. Contemporary Clinical Trials Communications
  16(February):100443, \doi{10.1016/j.conctc.2019.100443}

\bibitem[{Mikolov et~al.(2013)Mikolov, Sutskever, Chen, Corrado, and
  Dean}]{mikolov2013distributed}
Mikolov T, Sutskever I, Chen K, Corrado GS, Dean J (2013) Distributed
  representations of words and phrases and their compositionality. In: Advances
  in neural information processing systems, pp 3111--3119

\bibitem[{Minas et~al.(2018)Minas, Lagopoulos, and
  Tsoumakas}]{minas2018aristotle}
Minas A, Lagopoulos A, Tsoumakas G (2018) Aristotle university's approach to
  the technologically assisted reviews in empirical medicine task of the 2018
  clef ehealth lab. In: CLEF (Working Notes)

\bibitem[{Miwa et~al.(2014)Miwa, Thomas, O’Mara-Eves, and
  Ananiadou}]{miwa2014reducing}
Miwa M, Thomas J, O’Mara-Eves A, Ananiadou S (2014) Reducing systematic
  review workload through certainty-based screening. Journal of biomedical
  informatics 51:242--253

\bibitem[{Norman et~al.(2017)Norman, Leeflang, and
  N{\'{e}}v{\'{e}}ol}]{Norman2017}
Norman C, Leeflang M, N{\'{e}}v{\'{e}}ol A (2017) {LIMSI@CLEF eHealth 2017 task
  2: Logistic regression for automatic article ranking}. In: CEUR Workshop
  Proceedings, vol 1866

\bibitem[{Norman et~al.(2018)Norman, Leeflang, and
  N{\'{e}}v{\'{e}}ol}]{Norman2018}
Norman C, Leeflang M, N{\'{e}}v{\'{e}}ol A (2018) {LIMSI@CLEF eHealth 2018 Task
  2: Technology assisted reviews by stacking active and static learning}. In:
  CEUR Workshop Proceedings, vol 2125

\bibitem[{Nunzio(2019)}]{Nunzio2019}
Nunzio GMD (2019) {A distributed effort approach for systematic reviews. IMS
  UNIPD at CLEF 2019 EHealth Task 2}. In: CEUR Workshop Proceedings, vol 2380

\bibitem[{Pagliardini et~al.(2018)Pagliardini, Gupta, and Jaggi}]{pgj2017unsup}
Pagliardini M, Gupta P, Jaggi M (2018) {Unsupervised Learning of Sentence
  Embeddings using Compositional n-Gram Features}. In: NAACL 2018 - Conference
  of the North American Chapter of the Association for Computational
  Linguistics

\bibitem[{Pavlopoulos et~al.(2014)Pavlopoulos, Kosmopoulos, and
  Androutsopoulos}]{pavlopoulos2014continuous}
Pavlopoulos I, Kosmopoulos A, Androutsopoulos I (2014) Continuous space word
  vectors obtained by applying word2vec to abstracts of biomedical articles.
  Word Journal Of The International Linguistic Association pp 1--4

\bibitem[{Qin et~al.(2010)Qin, Liu, Xu, and Li}]{qin2010letor}
Qin T, Liu TY, Xu J, Li H (2010) Letor: A benchmark collection for research on
  learning to rank for information retrieval. Information Retrieval
  13(4):346--374

\bibitem[{Robertson(2010)}]{Robertson2010}
Robertson S (2010) {The Probabilistic Relevance Framework: BM25 and Beyond}.
  Foundations and Trends{\textregistered} in Information Retrieval
  3(4):333--389

\bibitem[{Sackett(1997)}]{sackett1997evidence}
Sackett DL (1997) Evidence-based medicine. In: Seminars in perinatology,
  Elsevier, vol~21, pp 3--5

\bibitem[{Scells et~al.(2017)Scells, Zuccon, Deacon, and Koopman}]{Scells2017}
Scells H, Zuccon G, Deacon A, Koopman B (2017) {QUT ielab at CLEF eHealth 2017
  technology assisted reviews track: Initial experiments with learning to
  rank}. In: CEUR Workshop Proceedings, vol 1866

\bibitem[{Scells et~al.(2020{\natexlab{a}})Scells, Zuccon, and
  Koopman}]{Scells2020OldDog}
Scells H, Zuccon G, Koopman B (2020{\natexlab{a}}) {You Can Teach an Old Dog
  New Tricks: Rank Fusion applied to Coordination Level Matching for Ranking in
  Systematic Reviews}. In: Lecture Notes in Computer Science (including
  subseries Lecture Notes in Artificial Intelligence and Lecture Notes in
  Bioinformatics), Springer International Publishing, vol 12035 LNCS, pp
  399--414

\bibitem[{Scells et~al.(2020{\natexlab{b}})Scells, Zuccon, Koopman, and
  Clark}]{Scells2020}
Scells H, Zuccon G, Koopman B, Clark J (2020{\natexlab{b}}) {A Computational
  Approach for Objectively Derived Systematic Review Search Strategies}. In:
  Lecture Notes in Computer Science (including subseries Lecture Notes in
  Artificial Intelligence and Lecture Notes in Bioinformatics), Springer
  International Publishing, vol 12035 LNCS, pp 385--398

\bibitem[{Singh et~al.(2017)Singh, Marshall, Thomas, and Wallace}]{Singh2017}
Singh G, Marshall I, Thomas J, Wallace B (2017) {Identifying diagnostic test
  accuracy publications using a deep model}. In: CEUR Workshop Proceedings, vol
  1866

\bibitem[{Tsafnat et~al.(2014)Tsafnat, Glasziou, Choong, Dunn, Galgani, and
  Coiera}]{tsafnat2014systematic}
Tsafnat G, Glasziou P, Choong MK, Dunn A, Galgani F, Coiera E (2014) Systematic
  review automation technologies. Systematic reviews 3(1):74

\bibitem[{{Van Altena} and Olabarriaga(2017)}]{VanAltena2017}
{Van Altena} AJ, Olabarriaga SD (2017) {Predicting publication inclusion for
  diagnostic accuracy test reviews using random forests and topic modelling}.
  In: CEUR Workshop Proceedings, vol 1866

\bibitem[{Wallace et~al.(2010{\natexlab{a}})Wallace, Small, Brodley, and
  Trikalinos}]{wallace2010active}
Wallace BC, Small K, Brodley CE, Trikalinos TA (2010{\natexlab{a}}) Active
  learning for biomedical citation screening. In: Proceedings of the 16th ACM
  SIGKDD international conference on Knowledge discovery and data mining, pp
  173--182

\bibitem[{Wallace et~al.(2010{\natexlab{b}})Wallace, Trikalinos, Lau, Brodley,
  and Schmid}]{wallace2010semi}
Wallace BC, Trikalinos TA, Lau J, Brodley C, Schmid CH (2010{\natexlab{b}})
  Semi-automated screening of biomedical citations for systematic reviews. BMC
  bioinformatics 11(1):1--11

\bibitem[{Wu et~al.(2018)Wu, Wang, Chen, Chen, Hu, and He}]{Wu2018}
Wu H, Wang T, Chen J, Chen S, Hu Q, He L (2018) {ECNU at 2018 eHealth Task 2:
  Technologically assisted reviews in empirical medicine}. In: CEUR Workshop
  Proceedings, vol 2125

\bibitem[{Yu and Menzies(2017)}]{Yu2017}
Yu Z, Menzies T (2017) {Data balancing for technologically assisted reviews:
  Undersampling or reweighting}. In: CEUR Workshop Proceedings, vol 1866

\bibitem[{Zou et~al.(2018)Zou, Li, and Kanoulas}]{Zou2018}
Zou J, Li D, Kanoulas E (2018) {Technology assisted reviews: Finding the last
  few relevant documents by asking yes/no questions to reviewers}. 41st
  International ACM SIGIR Conference on Research and Development in Information
  Retrieval, SIGIR 2018 pp 949--952

\bibitem[{Zou and Kanoulas(2020)}]{Zou2020}
Zou JIE, Kanoulas E (2020) {Towards Question-based High-recall Information
  Retrieval: Locating the Last Few Relevant Documents for Technology-assisted
  Reviews}. ACM Transactions on Information Systems 38(3)

\end{thebibliography}


\end{document}